\documentclass[hyper,11pt,letterpaper]{JHEP3}
\usepackage[dvips]{epsfig}
\usepackage{amsmath,amssymb,epsf,amsfonts}
\usepackage{graphicx}
\usepackage{dcolumn}
\usepackage{bm}

\usepackage[english]{babel}
\usepackage{amssymb}
\usepackage{amsfonts}
\usepackage{amsmath}
\usepackage{graphicx}
\usepackage{epsfig}
\usepackage{cite}
  \newcommand{\cF}{{\cal F}}
  
  \newcommand{\cJ}{{\cal J}}
  \newcommand{\cL}{{\cal L}}
  \newcommand{\cN}{{\cal N}}

  \newcommand{\bbR}{{\mathbb R}}

\newcommand{\be}{\begin{equation}} \newcommand{\ee}{\end{equation}}
\newcommand{\bea}{\begin{eqnarray}} \newcommand{\eea}{\end{eqnarray}}
\newcommand{\beann}{\begin{eqnarray*}}  \newcommand{\eeann}{\end{eqnarray*}}
\newcommand{\bfig}{\begin{figure}} \newcommand{\efig}{\end{figure}}
\newcommand{\ba}{\begin{array}} \newcommand{\ea}{\end{array}}
\newcommand{\bcen}{\begin{center}} \newcommand{\ecen}{\end{center}}
\newcommand{\btab}{\begin{tabular}} \newcommand{\etab}{\end{tabular}}

\def\tr{\operatorname{tr\:}}     
   \def\dim{\operatorname{dim}}

\renewcommand{\Re}{\mathop{\rm Re}}   
  \def\Nfour{{\cal N}=4 }

\newtheorem{Proposition}{Proposition}[section]

\newtheorem{Theorem}{Theorem}[section]
\newtheorem{Lemma}{Lemma}[section]
\newtheorem{Corrolary}{Corrolary}[section]

\newcommand{\bp}{\begin{Proposition}}   \newcommand{\ep}{\end{Proposition}}
\newcommand{\bt}{\begin{Theorem}}   \newcommand{\et}{\end{Theorem}}
\newcommand{\bl}{\begin{Lemma}}     \newcommand{\el}{\end{Lemma}}
\newcommand{\bc}{\begin{Corrolary}} \newcommand{\ec}{\end{Corrolary}}

\def\gb{\bar{g}}
\def\xb{\underline x}
\def\ub{\underline u}
\def\rb{\underline r}

\title{Minimal area submanifolds in AdS $\times$ compact}

\author{C. Robin Graham\footnotemark[1]\,
and
Andreas Karch\footnotemark[2]
\\
\footnotemark[1] \,: Department of Mathematics, University of Washington, Seattle, WA
98195-4350, \\
E-mail: \email{robin@math.washington.edu}
\\
\footnotemark[2] \,: Department of Physics, University of Washington, Seattle, WA
98195-1560, \\
E-mail: \email{karch@phys.washington.edu}
}

\abstract{We describe the asymptotic behavior of minimal area submanifolds
in product spacetimes of an asymptotically hyperbolic space times
a compact internal manifold. In particular, we find that unlike the case of
a minimal
area submanifold just in an asymptotically hyperbolic space,
the internal part of the boundary submanifold is constrained to be itself a
minimal area submanifold. For applications to holography, this tells us
what are the allowed ``flavor branes" that can be added to a holographic
field theory. We also give a compact geometric expression for the spectrum
of operator dimensions associated with the slipping modes of the
submanifold in the internal space. We illustrate our results with several
examples, including some that haven't appeared in the literature before.}

\begin{document}

\section{Introduction}\label{intro}

The gauge/gravity correspondence
\cite{Maldacena:1997re,Gubser:1998bc,Witten:1998qj}
or ``holography" for short, provides
a large class of solvable models of strong coupling dynamics. These solvable
toy models are being employed to understand qualitative aspects
of an ever growing array of physics questions, spanning from nuclear to atomic
and condensed matter physics.
The mathematics underlying holographic techniques is 28 years
old
and predates its application to physics \cite{Fefferman:1985}.
Holography relates
strongly coupled quantum field theories in $n$ spacetime dimensions
to a gravitational problem in one higher
dimension. The solution of the system via holography therefore becomes
essentially a geometric problem: the geometry of the extra dimension
encodes all the properties of the strongly coupled physical system. In the
simplest cases, the higher dimensional space needs to be of a very special
geometric type: a product manifold with one
factor that is an asymptotically hyperbolic Einstein space
and the other factor a compact ``internal" space.
In essence,
the problem is one of constructing solutions to a non-linear second order
partial differential equation (encoding the Einstein condition) subject to
certain asymptotic boundary conditions.
The asymptotic behavior of such spaces at infinity has been described in
in \cite{Fefferman:1985,deHaro:2000xn,Fefferman:2012}.

There are several questions within these holographic toy models for
which one, in addition to the background geometry, is also interested
in finding a minimal\footnote{Following the tradition in the mathematical
  literature, by ``minimal" area submanifold we simply mean a submanifold
  whose area doesn't change at linear order under small fluctuations. They
  can be minima, maxima or saddles of the area functional.  A minimal
  submanifold is referred to as stable if the second order variation of area
  is positive definite.  In particular, stable minimal submanifolds are
  local minima of the area functional.  \label{minstab}} area submanifold within
that space. Let us briefly
recall three classes of questions that rely on minimal area submanifolds.

The first appearance of minimal area submanifolds in holography was in the context
of Wilson lines \cite{Maldacena:1998im,Rey:1998ik}. A Wilson line measures
the response of the strongly coupled system to inclusion of an external test
particle, following a predescribed worldline. To calculate the
expectation value of the operator describing this insertion, the holographic
recipe is to calculate the area of the minimal area surface in the bulk ending
on the worldline.

A second example of an application of minimal area submanifolds in
holography
is probe flavor branes\cite{Karch:2002sh}.
Flavor branes are needed to incorporate quarks into holographic
models of QCD; for condensed matter applications they can introduce
the charge carriers (the electrons) into a strongly coupled
phonon bath modeled, for example, by $\Nfour$ Super-Yang Mills (SYM).
The extra degrees of freedom added to the quantum field
theory can either live in the whole $n$ dimensional field theory spacetime,
or only on a $k$ dimensional subspace (a ``defect"). The holographic description
requires a minimal area submanifold ending on the location of the defect.
In addition, different flavor branes wrap different submanifolds of
the internal space, corresponding to different matter content and interactions
of the extra degrees of freedom added to the field theory. The worldvolume
of the flavor brane only has to be a minimal area submanifold when regarded
as a submanifold of the full product spacetime that constitutes the
holographic
dual, not separately as a submanifold of the hyperbolic and the compact
factor. The interplay of the shape of the submanifold in the internal and
hyperbolic factor are crucial to
model even the simplest physical parameters such a theory should have, for
example the mass of the extra fields. The Wilson line can be viewed
as a special case of a flavor brane, where the matter added is a 0+1
dimensional defect, with the degree of freedom living on it being the
external test quark.

The reason that in these cases the original geometric question of finding
an Einstein manifold gets replaced by a minimal area problem is the
``probe approximation" inherent in this construction. In general, changing
the theory by adding extra matter would require to re-solve
the system of coupled differential equations describing the bulk
geometry. The tension of the flavor brane gives rise to a non-zero stress
tensor that appears as a source on the right-hand side of the bulk Einstein
equations. The worldvolume of the brane however wants to minimize its area
in this backreacted geometry. This way one finds a new background for the
new field theory. In the limit that the extra degrees of freedom added are
much fewer than the degrees of freedom in the
original theory\footnote{
In cases where the strongly coupled field theory is a non-Abelian
gauge theory with $SU(N)$ gauge group,
this is typically the case when the added matter
is in the fundamental representation of the gauge group. While there
are of the order $N^2$ gauge fields,
there are only of order $N$ matter fields.}, the stress tensor associated
to the extra matter is negligible and
consequently it does not
backreact on the geometry. The flavor brane simply minimizes its own
worldvolume area in a fixed background geometry.

The third and most recent example of studying minimal area
problems in holography is entanglement entropies. While we will keep our
discussion in the Euclidean setting unless
explicitly mentioned otherwise, entanglement entropy is intrinsically tied
to Lorentzian signature. If at a given time $t_0$ the $n-1$ dimensional
field theory space is separated
into two regions by a $n-2$ dimensional surface,  
one can, at that instant, associate an entropy to the field theory living in
one of the regions by tracing over the degrees of freedom in the other region.
This is called the entanglement entropy.
The holographic proposal of \cite{Ryu:2006bv} demands that the
corresponding entanglement entropy is given by the area of a minimal
area submanifold ending on the surface dividing the two regions, measured in
Planck units (that is $S=A/(4 G)$, where $S$ is the entanglement entropy,
$A$ the area and $G$ Newton's constant in the bulk).
For the special case that the bulk is static, for example if it is $n+1$
dimensional Anti de-Sitter space (AdS$_{n+1}$), the minimal area in the
bulk will also be at the same fixed time $t=t_0$; the problem then reduces
to finding a minimal area submanifold in the $t=t_0$ submanifold of
AdS$_{n+1}$, which itself is an $n$ dimensional hyperbolic space.
Note that in this
proposal the bulk minimal area should be co-dimension 2, just as the field
theory surface it ends on.
Consequently, in the case where the bulk
is a product manifold this minimal area submanifold always wraps
the entire internal space. Unlike the case of flavor branes, here the
problem essentially reduces to that of finding minimal area submanifolds
in the asymptotically hyperbolic Einstein space itself.

The goal of our work is to fully
describe the asymptotic structure of the most general minimal area submanifold
in asymptotically hyperbolic
Einstein spaces times a compact internal manifold which itself is
asymptotic to a product of a submanifold in each factor. The corresponding
problem in an asymptotically hyperbolic Einstein space itself has been
solved in \cite{Graham:1999pm}. In particular for the case of
flavor branes, it is crucial to address the question to what
extent this picture changes when one asks for minimal areas in the
product space-time. The resulting structure is indeed much richer than what
one gets simply in an asymptotically hyperbolic Einstein space.

This paper is mostly dedicated to a physics audience.  We derive two main
results.  We find an interesting new constraint and we give an explicit
geometric formula for the operator dimensions of operators dual to
deformations of the cycle.  In a more mathematical companion paper we will
analyze the general formal asymptotics of minimal area submanifolds
asymptotic to products.

Our constraint restricts the asymptotic form the submanifold
can have.  While the submanifold occupied by
a defect flavor brane in the field theory dimensions is arbitrary,
the asymptotic submanifold it wraps in the internal space
has to be minimal itself. This constrains
the form of potential flavor branes one may wish to add to the geometry.
For the operator spectrum, we consider submanifolds that can be viewed as a
first order deformation of a product submanifold, with the requirement that
the deformation vanishes asymptotically.  According to the AdS/CFT
dictionary, these dimensions correspond
in the bulk to the indicial roots for the linearized minimal submanifold
equations.  We show that the dimensions are
governed by two geometric operators defined by the geometry of the minimal
submanifold in the internal space: the scalar Laplace operator and the
Jacobi operator.  The derivations of both the minimality constraint and the 
operator spectrum only require the spacetime background metric to be
asymptotically hyperbolic; they do not use the Einstein condition.
We also work out several novel examples of minimal area
submanifolds; in particular we give examples of submanifolds where
perturbations of the product submanifold are turned on in both factors
simultaneously.

The organization of this paper is as follows: in the next section we
give the description of the full asymptotic data starting
from the case of a product submanifold and working out its deformations.
In section 3 we derive the constraint that the boundary internal
submanifold must be minimal.
In section 4 we give several examples as well as their physical
interpretation.

An important aspect of the minimal area problem is
the calculation of the renormalized area.  Due to the singularity of the
metric near the
boundary of asymptotically hyperbolic spaces, the area of all the
submanifolds we consider is divergent. To assign a finite answer
for the actual area one has to carefully understand the various divergent
contributions to the area and cancel them with appropriate local
counterterms. In the physics literature this is known as ``holo-RG". 
The procedure is well-understood for minimal area submanifolds of 
asymptotically hyperbolic Einstein spaces.  When internal variables are  
allowed, the program has been carried out for special examples in
\cite{Karch:2005ms}.  We leave for the future the study of the 
renormalized area for the general minimal area problem in spaces with
internal variables.    

\subsection*{Notation:}

We consider background spacetimes of the form $X^{n+1}\times K^m$,
where $X^{n+1}$ is a $n+1$ dimensional manifold with asymptotically
hyperbolic metric $g_X$ and boundary
$\partial X=M$ and $K^m$ is a compact
$m$ dimensional manifold with metric $g_K$.  We take the metric on
$X^{n+1}\times K^m$ to be the product
$$
g=g_X + g_K
$$
and we work in Euclidean signature.  The assumption that $g_X$ is
asymptotically hyperbolic means that in an appropriate choice of
coordinates it can be written in the form
\be\label{gX}
g_X=\frac{dr^2 + \bar{g}_r}{r^2}
\ee
where $\partial X=\{r=0\}$ and $\bar{g}_r$ is a 1-parameter family of
metrics on $M$ smooth in $r$ up to $r=0$.

Let us choose submanifolds $N^k\subset
M$ and $\Sigma^l\subset K$.
We are interested in minimal submanifolds $Z \subset X \times K$ with
$\partial Z=N \times \Sigma \subset M \times K$.  For simplicity, we take all
our submanifolds to be embedded (no self-intersections) and regular at
infinity.  We can
choose local coordinates $(x^{\alpha},u^{\alpha'})$ on $M$ and
$(s^A,t^{A'})$ on $K$ in such a way that $N\subset M$ is
given by $N=\{u^{\alpha'}=0\}$ and $\Sigma\subset K$ is given by
$\Sigma=\{t^{A'}=0\}$,
and further
\be\label{orthog}
 \left. \bar{g}_{\alpha \beta'} \right |_{r=0,u=0} =\left .  g_{A B'}
\right |_{t=0} =0 .\ee
That is, the $t$ and $u$ coordinates on the boundary are ``orthogonal" to
the defect.  In such coordinates in which \eqref{gX} also holds, our
minimal submanifold $Z$ is described by giving the $u$'s and $t$'s as
functions of $(x,r,s)$, and the boundary condition reads $u=0$, $t=0$ at
$r=0$.  

\section{Product submanifolds and their deformations}
\subsection{Product submanifolds}\label{prod}

The simplest examples of minimal area submanifolds in asymptotically AdS
times compact product spaces are submanifolds where not just the boundary
of $Z$ is a
product manifold $\partial Z = N \times \Sigma$, but where $Z$ itself is a
product $Z_0=N' \times \Sigma$ with $\partial N'=N$.
In fact, any product of the form $N' \times \Sigma$ is
a minimal area submanifold in $X \times K$  as long as
 $N'$ is a
minimal area submanifold in $X$ and $\Sigma$ a minimal area submanifold in $K$.
The converse is true too: if $N'\times \Sigma$ is minimal in $X \times K$, then
$N'$ must be minimal in $X$ and $\Sigma$ minimal in $K$.

A general
analysis of the formal asymptotics of minimal submanifolds $N'$ of $X$ has
been performed already in \cite{Graham:1999pm}.
In this case, the local data
that needs to be specified near the boundary is an arbitrary submanifold
$\partial N' =N$ of the boundary manifold $M$. The shape of $N$ is
completely unconstrained.  A minimal submanifold $N'$ necessarily
intersects $\partial X$ orthogonally.  One describes $N'$ by giving
$u^{\alpha'}$ as a function of $x$ and $r$.  $u^{\alpha'}(x,r)$ is then
specified as a power series in $r$ (and its logarithm). The
coefficient of $r^{k+2}$ is locally undetermined.  It is typically fixed by
some global requirements, such as smoothness, on the submanifold.  Also there can be
different minimal submanifolds $N'$ with the same boundary
submanifold $N$ and these will typically have different values for the
$r^{k+2}$ coefficient.  Once the coefficient of
$r^{k+2}$ is fixed, there is no more freedom in the series expansion
defining $u(x,r)$.  In stark contrast, the internal factor $\Sigma$ already
needs to be minimal in $K$ to begin with. So for the case of product
submanifolds, we say that the {\it boundary data for the minimal product
  submanifold $Z_0$ consists of an arbitrary submanifold $N$ of $M$
  together with a {\it minimal} submanifold of $K$.}

In the following we want to analyze the generic structure of minimal
submanifolds $Z$ that are obtained by infinitesimal perturbations of a
product minimal submanifold $Z_0=N'\times \Sigma$. The requirement that our
submanifold has as its boundary a
product $N \times \Sigma$ means that we require that the perturbation goes
to zero at the boundary; it need not be small in the bulk of the
space and need not itself be of product form. In the remainder of this
section we will identify
the complete local data that determines such an infinitesimally perturbed
product submanifold. Elsewhere we will show that
relaxing the requirement that the submanifold can be written as the
deformation of a product does not lead to any additional freedom.

\subsection{Jacobi operator}

The spectrum of small fluctuations around a minimal submanifold is
governed by the Jacobi operator ${\cJ}$ of the submanifold.  We digress
briefly to review the Jacobi operator since it plays a central role in our
discussion.  A more detailed discussion can be found, for instance, in
\cite{Lawson:1980ms}.

In general, suppose that $Z$ is a submanifold of a space $X$ with a
Riemannian metric $g_X$.  Perturbations of $Z$ can be described
by 1-parameter families of maps $F_t:Z\rightarrow Z_t\subset X$ for small
$t$ satisfying $F_0=$ Identity.  The derivative
$\delta Z=\partial_tF_t|_{t=0}$ can be interpreted as a vector field
defined on $Z$, which we assume is everywhere normal to $Z$ since
tangential vector fields correspond to reparametrizations of $Z$.
The condition
that $Z$ is minimal is the requirement that $\partial_tA(Z_t)|_{t=0}=0$
for all maps $F_t$, where $A$ denotes area.  This can be expressed by the
vanishing of the mean curvature vector of $Z$.  Recall that the mean
curvature vector is defined as follows.  The second
fundamental form, or extrinsic curvature, $\cF$ of a submanifold $Z\subset
X$ is the symmetric
quadratic form on $TZ$ with values in the normal bundle $\cN$ given
by  $\cF(V,W)=({}^X\nabla_VW)^\perp$.  Here ${}^X\nabla$
denotes the Levi-Civita connection of $g_X$ and ${}^\perp$ the component
normal to $Z$.  The mean curvature vector $H$ of $Z$ is the trace of the
second fundamental form with respect to the induced metric on $Z$:
$H=\tr_{g_Z} \cF$.  So $H$ is a section of $\cN$ on $Z$.  Then
$$
\partial_tA(Z_t)|_{t=0}=-\int_Z\langle H, \delta Z\rangle \,dv_{g_Z},
$$
so $Z$ is minimal is the same as $H=0$.

Suppose now that $Z$ is minimal.  The second derivative of the area
function can be expressed as
$$
\partial_t^2A(Z_t)|_{t=0}= \int_Z\langle \cJ \delta Z,\delta Z\rangle \,
dv_{g_Z}.
$$
Here $\cJ$ is the Jacobi operator of $Z$, a differential operator acting on
sections of $\cN$.  The Jacobi operator can be expressed in invariant terms
as follows:
\begin{equation}\label{geometricJ}
\cJ=\nabla^*\nabla -\mathcal{R}-\mathcal{F}^2.
\end{equation}
We explain each of the three summands.  The first, $\nabla^*\nabla$, is the
normal bundle Laplacian.
The Levi-Civita connection ${}^X\nabla$ of $g_X$ induces a connection
$\nabla$ on $\cN$ defined by
$\nabla_VU=({}^X\nabla_V U)^\perp$.
Viewing $\nabla:\Gamma(\mathcal{N})\rightarrow
\Gamma(\mathcal{N}\otimes T^*Z)$, $\nabla^*$ denotes the adjoint
operator and $\nabla^*\nabla$ their composition.
Alternately, $\nabla^*\nabla$ may be expressed as
$$
(\nabla^*\nabla U)^{a'} = -g^{ab}\nabla_a\nabla_b U^{a'}.
$$
Unprimed indices correspond to tangent directions to $Z$ and primed
indices to normal directions.  On the right-hand side, $\nabla_a$ denotes
the normal bundle
connection coupled with the connection on $TZ$ induced by the Levi-Civita
connection on $X$.  The second term $\mathcal{R}$ in \eqref{geometricJ} is
a zeroth order term;
it is the linear transformation of the normal space at each point given by
$$
(\mathcal{R}U)^{a'}=R^{aa'}{}_{ab'}U^{b'},
$$
where $R$ denotes the curvature tensor of the background metric $g_X$.
$\mathcal{R}$ is perhaps best viewed
as a partial mixed version of the Ricci tensor:  it is the
normal part of the tangential trace of the curvature tensor, viewed as a
linear transformation of $\cN$.  The third term $\cF^2$ in
\eqref{geometricJ} is another zeroth order term; it is a linear
transformation of the normal space which is quadratic in $\cF$.  Its action
on a normal vector $U^{a'}$ is given by
$$
(\mathcal{F}^2U)^{a'}=\mathcal{F}^{a'}_{ab}\mathcal{F}_{b'}^{ab}U^{b'}. 
$$
So $\mathcal{F}^2$ is the norm-squared of $\mathcal{F}$ in the
tangential indices, viewed as a linear transformation in the normal
indices.  At each point it is a positive semi-definite transformation of
$\cN$.

We apply this discussion to our product situation, taking the background
space to be $X\times K$ with its product metric $g_X + g_K$, and taking
the minimal submanifold to be $Z_0=N'\times \Sigma$.  Our
perturbation $\delta Z$ is required to satisfy the linearized minimal
submanifold equation, which is to say that it must be in the kernel of the
Jacobi operator of $Z_0$.  Corresponding to the product decomposition of
our background space, we may write $\delta Z=((\delta Z)_X,(\delta Z)_K)$,
where $(\delta Z)_X$ is a normal vector to $N'$ (depending on both the
point in $N'$ and the point in $\Sigma$) and $(\delta Z)_K$ is a normal vector
to $\Sigma$ (depending on both the point in $\Sigma$ and the point in $N'$).  The
Jacobi operator of a product minimal submanifold itself
has a product decomposition:
\be\label{Jprod}
\cJ_{Z_0}(\delta Z)
=((\cJ_{N'}+\cL_\Sigma)(\delta Z)_X,(\cJ_{\Sigma}+\cL_{N'})(\delta Z)_K),
\ee
where $\cL=-\nabla^2$ denotes the scalar Laplacian on the
indicated space and $\cJ$ the Jacobi
operator of the indicated minimal submanifold.  The operators $\cL_\Sigma$
and $\cJ_\Sigma$ are self-adjoint elliptic operators on a compact manifold, so
each of them has a spectral decomposition with eigenvalues going
to $+\infty$.
Upon diagonalizing these operators, it is evident that the behavior of
$\delta Z$ near the boundary is determined by these eigenvalues and by the
form of $\cL_{N'}$ and $\cJ_{N'}$ near the boundary.

\subsection{Fluctuation in the internal space}\label{intfluc}
First consider the behavior of $(\delta Z)_K$.
The equation of motion which follows from \eqref{Jprod} is
$$
\left (\cL_{N'} + \lambda) \right ) (\delta Z)_K =0
$$
where $\lambda$ is an eigenvalue for $\cJ_\Sigma$.
This is exactly the equation for a massive scalar field on the space $N'$
with mass squared given by $\lambda$.
Since $N'$ intersects the boundary orthogonally, the induced metric is
asymptotically hyperbolic, so the near boundary behavior is
given by $r^{\Delta_{\pm}}$ with the standard mass/dimension relation
\be
\label{deltainternal}
 \Delta_{\pm} = \frac{k}{2} \pm \frac{ \sqrt{k^2 + 4 \lambda}}{2}. \ee
Our boundary condition requires\footnote{In the holographically dual field
theory excitations with real and negative $\Delta$ correspond to irrelevant operators. These
can not be added to the action as they would spoil the short distance
properties of the field theory. We can
however add them as sources with delta-function support and calculate
correlation functions.} $\operatorname{Re}\Delta>0$, so only
negative $\lambda$
produce a valid $\Delta_-$.  There are at most finitely many such
$\lambda$, corresponding to the perturbations of $\Sigma$ in directions for
which the area decreases.
In terms of the holographically dual field theory these fluctuations map to
the relevant operators, that is operators whose effect becomes negligible
at short distances.  When
$\lambda$ is very negative; namely $\lambda<-k^2/4$, $\Delta_{\pm}$ are a
complex conjugate pair with real part $k/2$, corresponding to a
scattering phenomenon (when viewing hyperbolic space as the target space of
particle motion) or equivalently to an instability (when studying
Lorentzian AdS).  For $-k^2/4<\lambda<0$, $\Delta_\pm$ are real with
$0<\Delta_-<k/2$, $k/2<\Delta_+<k$.  Even though the coefficients of both
$r^{\Delta_-}$ and $r^{\Delta_+}$ are formally undetermined as functions on
$N$, one anticipates that generically
the coefficient of $r^{\Delta_-}$ (if $\lambda<0$) can be chosen
arbitrarily but that the coefficients of all of the $r^{\Delta_+}$ will be
determined by global considerations.  If $\lambda =0$, then $\Delta_-=0$
which violates our boundary condition that $Z$ asymptotically approaches
$N\times \Sigma$.  However, $\lambda=0$ corresponds to a minimal perturbation of
$\Sigma\subset K$, so corresponds to a deformation of $Z_0$ in which $\Sigma$ changes
and the perturbations $Z$ remain products. In the holographically dual
field theory, this maps to what is called a marginal operator.

Negative eigenvalues for $\cJ_\Sigma$ thus play an important role in our
analysis because
they correspond to additional freedom to prescribe local boundary
conditions for minimal submanifolds $Z$.
Clearly $\cJ_\Sigma$ depends solely on the geometry of $\Sigma\subset K$,
i.e. it is independent of the asymptotically AdS space and its
submanifold.  Thus the same is true of the $\lambda$'s.  The AdS geometry
influences the $\Delta$'s only through $k$, the dimension of the AdS
boundary submanifold.  Note that $\cJ$ scales like
$(distance)^{-2}$, so the $\lambda$'s scale the same way.  In particular,
the number of negative $\lambda$'s is independent of rescaling $g_K$.
Since
$\nabla^*\nabla\geq 0$, negative $\lambda$'s must be created by the
influence of $\mathcal{R}$ and $\mathcal{F}^2$.  Since $\mathcal{F}^2\geq
0$, this term always has a negative effect on $\lambda$.  The
$\mathcal{R}$ term has a
negative effect for manifolds of positive sectional curvature and vice
versa.
 One
can easily read off some qualitative information from such considerations.
For example, if $K$ has non-positive
sectional curvature, say a compact hyperbolic manifold or a torus, and $\Sigma$
is totally geodesic (i.e. $\cF=0$), or even just has
sufficiently small extrinsic curvature in the case that $K$ has negative
sectional curvature, then $\cJ_\Sigma\geq 0$ so there are no positive
$\Delta_-$'s and no locally prescribable freedom in the expansion of
$(\delta Z)_K$.  On the other hand,
if $K$ has positive curvature and $\Sigma$ is totally geodesic, then we
anticipate the possibility of negative $\lambda$'s and therefore additional
freedom for minimal
submanifolds $Z$.  This freedom will be exhibited in \S\ref{examples} when
$\Sigma$ is an equatorial sphere embedded in a higher-dimensional sphere.

In conclusion, we find that {\it for every eigenvector with negative
  eigenvalue of the Jacobi operator ${\cal J}_\Sigma$ associated to the
  embedding of   $\Sigma$ in $K$, there is one piece of local information
  that needs to be
  specified at the boundary: the coefficient of $r^{\Delta_-}$ (as a
  function on $N$). For every eigenvector of the Jacobi operator (even with
  positive eigenvalue), the coefficient of $r^{\Delta_+}$ is undetermined
  by   the boundary data and needs to be specified by global
  considerations. $\Delta_{\pm}$ are given by \eqref{deltainternal}.}

Fluctuations of the brane which correspond to eigenfunctions
of ${\cal J}_\Sigma$ with positive eigenvalue $\lambda$ encode the spectrum of
an infinite tower of relevant operators in the dual field theory whose
dimensions
are given by $\Delta_+$ from \eqref{deltainternal}.

\subsection{Fluctuation in AdS}

Now consider the behavior of $(\delta Z)_X$.  The equation of motion
reads
$$
(\cJ_{N'}+\Lambda)(\delta Z)_X=0
$$
where $\Lambda$ is an eigenvalue of $\cL_\Sigma$.  We need to determine
the leading term in $\cJ_{N'}(\delta Z)_X$ under the power law ansatz
$(\delta Z)_X\sim r^{\Delta}$.  Choose coordinates
$(x^\alpha,u^{\alpha'},r)$ on $X$ so that $u^{\alpha'}$ vanishes on $N'$
and $r$ vanishes on $\partial X$.
($g_X$ need not have the form \eqref{gX} in these coordinates.)
We use a $0$ index to
correspond to $r$ and let $\mu$, $\nu$, $\sigma$ run over $\alpha$ and $0$
and $i$,
$j$, $k$ run over all of $\alpha$, $\alpha'$, $0$.  Let
$g$ denote the asymptotically hyperbolic metric $g_X$.  Using the fact that
$N'$ intersects $\partial X$ orthogonally, we can arrange
that $g^{\mu\alpha'}=0$ everywhere on $N'$, so that $\mu$ corresponds to
directions tangent to $N'$ and $\alpha'$ to directions normal to $N'$.
Write $g=r^{-2}\gb$.  The asymptotically hyperbolic condition says that
$\gb^{00}=1$ on the boundary.  The Christoffel symbols of $g$ satisfy
$$
\Gamma_{ij}^k\sim -r^{-1}(r_j\delta_i^k+r_i\delta_j^k -\gb^{kl}r_l\gb_{ij}).
$$
Thus for the $O(r^{-1})$ terms we have
\[
\begin{split}
\Gamma_{\mu\nu}^\sigma&\sim -r^{-1}(\delta_\mu^0\delta_\nu^\sigma
+\delta_\nu^0\delta_\mu^\sigma-\gb^{\sigma 0}\gb_{\mu\nu})\\
\Gamma_{\mu\nu}^{\alpha'}&\sim 0\\
\Gamma_{\mu\beta'}^{\alpha'}&\sim
-r^{-1}\delta_{\mu}^0\delta_{\beta'}^{\alpha'}.
\end{split}
\]
If $U^{\alpha'}$ is a vector field normal to $N'$ and smooth up to the
boundary, then
$$
\nabla_\nu (r^\Delta U^{\alpha'})
\sim \Delta r^{\Delta-1}\delta_\nu^0U^{\alpha'} +r^\Delta
\Gamma_{\nu\beta'}^{\alpha'}U^{\beta'}
\sim (\Delta -1)r^{\Delta -1}\delta_\nu^0U^{\alpha'}
$$
so
\[
\begin{split}
g^{\mu\nu}\nabla_\mu\nabla_\nu(r^\Delta& U^{\alpha'})
\sim (\Delta -1)g^{\mu\nu}\nabla_\mu(r^{\Delta -1}\delta_\nu^0U^{\alpha'})\\
&\sim (\Delta -1)g^{\mu\nu}
\left[(\Delta  -1)r^{\Delta-2}\delta_\mu^0\delta_\nu^0U^{\alpha'}
+r^{\Delta-1}\delta_\nu^0\Gamma_{\mu\beta'}^{\alpha'}U^{\beta'}
-r^{\Delta-1}\Gamma_{\mu\nu}^{\sigma}\delta_\sigma^0U^{\alpha'}\right]\\
&\sim (\Delta -1)g^{\mu\nu}r^{\Delta-2}
\left[(\Delta  -1)\delta_\mu^0\delta_\nu^0U^{\alpha'}
-\delta_\mu^0\delta_\nu^0U^{\alpha'}
+(\delta_\mu^0\delta_\nu^\sigma
+\delta_\nu^0\delta_\mu^\sigma-\gb^{\sigma
  0}\gb_{\mu\nu})\delta_\sigma^0U^{\alpha'}\right]\\
&\sim
(\Delta-1)r^{\Delta}\left[(\Delta-1)U^{\alpha'}-U^{\alpha'}+(1-k)U^{\alpha'}\right]
=(\Delta-1)(\Delta -k-1)U^{\alpha'}.
\end{split}
\]
The curvature tensor of $g$ satisfies
$$
R^{ij}{}_{kl}\sim -(\delta^i_k\delta^j_l-\delta^i_l\delta^j_k)
$$
so that
$$
R^{\mu \alpha'}{}_{\mu\beta'}\sim -(k+1)\delta^{\alpha'}_{\beta'},
$$
and we have $\cF^2\sim 0$.  Hence
$$
-\cJ_{N'}(r^{\Delta}U^{\alpha'})
\sim \left[(\Delta-1)(\Delta-k-1)-(k+1)\right]r^\Delta U^{\alpha'}
=\Delta(\Delta-k-2)r^\Delta U^{\alpha'}.
$$
We thus obtain $(\delta Z)_X\sim r^{\Delta_{\pm}}U^{\alpha'}$ with
\be
\label{deltaexternal}
\Delta_{\pm} = \frac{k+2}{2} \pm \frac{\sqrt{(k+2)^2 + 4 \Lambda}}{2}.
\ee
For the zero mode $\Lambda=0$ we recover the
result of \cite{Graham:1999pm} that $\Delta=k+2$ or $\Delta=0$.  Just as
for internal fluctuations, $\Delta=0$ corresponds to minimal perturbations
of
$N'$.  These always exist and there is no constraint on $U^{\alpha'}$ at
the boundary, corresponding to the freedom to choose the AdS boundary cycle
arbitrarily.  Since $\Lambda\geq 0$, no other $\Lambda$'s give a
$\Delta_-$ with $\Re \Delta_-\geq 0$, so we have no further freedom to
prescribe boundary data.  The coefficients of all $\Delta_+$ are formally
undetermined and need to be determined globally.

In conclusion, we find that {\it there is no local information to be
  specified except for the boundary submanifold $N$.  For every eigenvector
  of the scalar Laplacian on $\Sigma$, the coefficient of $r^{\Delta_+}$ is
  undetermined by the boundary data and needs to be specified by global
  considerations. $\Delta_{\pm}$ are given by \eqref{deltaexternal}.}

Fluctuations of the submanifold which correspond to
eigenfunctions of $\cL_\Sigma$ with positive eigenvalue $\Lambda$ encode
the spectrum of an infinite tower of relevant operators in the dual field
theory whose dimensions
are given by $\Delta_+$ from \eqref{deltaexternal}.

\section{Minimality constraint on internal submanifold}

We saw in \S\ref{prod} that in order that a product submanifold $N'\times
\Sigma$ be minimal, necessarily the internal factor $\Sigma$ must be minimal in $K$.
In this section we show that if $Z$ is a minimal submanifold of
$X\times K$, not necessarily a product, with $\partial Z=N\times \Sigma$, then
still $\Sigma$ must be minimal in $K$.  This result is global
in the internal submanifold $\Sigma$.  That is, we must assume that $\Sigma$ is a
full compact submanifold of $K$, as opposed to just a local piece
of one.  $Z$ only needs to exist near $N\times \Sigma$, but 
the result is false in general if $\Sigma$ is only a local piece of a
compact submanifold.

The outline of the argument is as follows.  Pick arbitrarily a point $p$ of
$N$ which remains fixed throughout.  For $s\in \Sigma$, the tangent
space to $Z$ at $(p,s)\in N\times \Sigma =\partial Z$ is a subspace of
$T_pX\times T_sK$.  Its projection to $T_pX$
can be written as a graph over $T_pN\times \mathbb{R}$, where the
$\mathbb{R}$ factor corresponds
to the $r$ variable.  The ``slope'' of this graph is a
vector $v\in \mathbb{R}^{n-k}$ which depends on $s$.  The fact
that $Z$ is minimal implies that $v(s)$ satisfies a nonlinear system of
partial differential equations as a vector-valued function on $\Sigma$.  A
direct analysis of this system of equations (integration by parts argument)
shows that the only global solution is $v=0$.  Geometrically this means
that $Z$ intersects the boundary orthogonally in the hyperbolic factor,
just like for product minimal submanifolds.  Finally, the fact that $v=0$
implies that $N'\times \Sigma$ is a product minimal submanifold in $T_p^+X\times
K$, where $T_p^+X$ denotes the interior half-space in $T_pX$ with its
induced hyperbolic metric, and $N'$ is a vertical plane in $T_p^+X$.  By
the result for product minimal submanifolds, we conclude that
$\Sigma$ must be minimal in $K$.

We proceed with the details.  Choose local coordinates
$(x,u,r,s,t)$ as described at the end of
\S\ref{intro}.  The $x$'s
restrict to a coordinate system on $N$ and
the $s$'s on $\Sigma$.  The $x$'s may be chosen so that the chosen point
$p$ has coordinates $x=0$.  $Z$ is given by
$$
Z=\{(x,u,r,s,t):u=\varphi(x,r,s),\,t=\psi(x,r,s)\},
$$
where $\varphi(x,0,s)=0$, $\psi(x,0,s)=0$.
Consider the submanifold $Z^\epsilon$ obtained
from $Z$ by dilating the $(x,u,r)$ coordinates in $X$ by $\epsilon$: 
$(\xb,\ub,\rb)=(\epsilon^{-1}x, \epsilon^{-1} u, \epsilon^{-1}r)$,
while leaving fixed the coordinates $(s,t)$ in $K$.  Thus
$$
Z^\epsilon=\{(\xb,\ub,\rb,s,t):\ub=\epsilon^{-1}\varphi(\epsilon
\xb,\epsilon \rb,
s),\,t=\psi(\epsilon \xb,\epsilon \rb,s)\}.
$$
If $g_X$ is given by
\eqref{gX} with $\bar{g}_r=\bar{g}(x,u,r)$, then in the dilated coordinates
it is given by $g^\epsilon_X=\rb^{-2}(d\rb^2 + \bar{g}^\epsilon_{\rb})$ with
$\bar{g}^\epsilon_{\rb}=\bar{g}(\epsilon \xb,\epsilon \ub,\epsilon \rb)$.
For each $\epsilon$, $Z^\epsilon$ is minimal with respect to
$g^\epsilon_X+g_K$.  Take the limit as $\epsilon\rightarrow 0$.  After
taking the limit, the variables $(\xb,\ub,\rb)$ should be regarded as
(infinitesimal) coordinates on the interior half-space $T_p^+X
=\{\rb>0\}$ in the tangent space $T_pX$.  In particular, they make sense
globally:  $\xb\in \mathbb{R}^k$, $\ub\in \mathbb{R}^{n-k}$, $\rb>0$.
In the limit we conclude that
\be\label{Z0}
Z^0 = \{(\xb,\ub,\rb,s,t): \ub= \rb v(s),\, t=0, \rb>0\}
\ee
is minimal for the metric
$g^0+g_K$, where $v(s)=\varphi_r (0,0,s)$ and
\be\label{g0}
g^0=\rb^{-2}(d\rb^2 + \bar{g}(0,0,0)).
\ee
Now $g^0$ is a constant curvature hyperbolic metric on $T_p^+X$.
We have $\partial Z^0=T_pN \times \Sigma$, where $T_pN=\{\ub=0,\,\rb=0\}$ is a
linear subspace in the boundary $\rb=0$ of our hyperbolic space.  For each
$s\in \Sigma$, the
projection of $Z^0$ into the hyperbolic factor $\{(\xb,\ub,\rb)\}$ is the
linear graph
$\ub=\rb v(s)$, $\rb>0$ in $T_p^+X$ with boundary $T_pN$.  This linear
graph varies with
$s$, and its ``slope'' $v(s)$ is a globally defined function on $\Sigma$ with
values in $\mathbb{R}^{n-k}$.

By direct analysis of the minimal submanifold
equations, we will show below that if $Z^0$ given by \eqref{Z0} is minimal
for $g^0+g_K$
and if $\Sigma\subset K$ is a compact submanifold, then necessarily $v=0$.  Our
desired conclusion follows immediately:  when $v=0$, $Z^0$ is a product
$Z^0=N'\times \Sigma$, where $N'=\{\ub=0\}$ is a vertical plane in hyperbolic
space.  As we saw in \S~\ref{prod}, this implies that $\Sigma$ is minimal
in $K$.  Since $v$ arose as $v=\varphi_r|_{r=0}$, the statement $v=0$ means
exactly that $Z$ intersects the boundary orthogonally in the hyperbolic
factor.

In \eqref{Z0}, $Z^0$ is expressed as a graph in the sense that the
$(\ub,t)$ variables are given as functions of the $(\xb,\rb,s)$ variables.
For simplicity, henceforth we remove the \underline{ } on
$(\xb,\ub,\rb)$, relabeling them $(x,u,r)$.
The minimal submanifold equation for a graph can be derived by calculating
the
Euler-Lagrange equation for the area functional.  This is carried out, e.g.,
in \S 2 of \cite{Graham:1999pm} for a graph expressed as $u=u(x)$ in a
space with coordinates $(x^\alpha,u^{\alpha'})$ and background metric $g$.
In our application, we will have to replace $x^\alpha$ by
$(x^\alpha,r,s^A)$ and $u^{\alpha'}$ by
$(u^{\alpha'},t^{A'})$.  For a graph $u=u(x)$, the induced metric $h$ is
given in the $x$ coordinates by
\begin{equation}\label{indmetric}
h_{\alpha\beta}=g_{\alpha\beta} + 2g_{\alpha'(\alpha}u^{\alpha'}_{,\beta)} +
g_{\alpha'\beta'}u^{\alpha'}_{,\alpha}u^{\beta'}_{,\beta},
\end{equation}
where the indices after a comma indicate coordinate differentiation.  The
minimal submanifold equation is:
\be\label{min}
\frac{1}{\sqrt {\det h}}\partial_\beta\left [ \sqrt{\det h} \,
    h^{\alpha\beta}
  \left ( g_{\alpha'\gamma'}u^{\alpha'}_{,\alpha}  +
  g_{\alpha\gamma'}\right ) \right ] -
  \frac12 h^{\alpha\beta} \left [ g_{\alpha\beta,\gamma'} +
  2g_{\alpha\alpha',\gamma'}u^{\alpha'}_{,\beta}
+ g_{\alpha'\beta',\gamma'}u^{\alpha'}_{,\alpha}u^{\beta'}_{,\beta} \right ] = 0.
\ee
Now $\alpha$ must be replaced by $(\alpha,r,A)$ and $\alpha'$ by
$(\alpha',A')$, and the background metric is $g=g^0 + g_K$, where $g^0$ is
given by \eqref{g0}.
By \eqref{orthog}, we can write
$\bar{g}(0,0,0)=\bar{g}_{\alpha\beta}dx^\alpha dx^\beta +
\bar{g}_{\alpha'\beta'}du^{\alpha'} du^{\beta'}$,
where $\bar{g}_{\alpha\beta}$ and $\bar{g}_{\alpha'\beta'}$ are constant.
Equation \eqref{min} breaks into two sets of equations corresponding to the
decomposition of the primed variables as $(\gamma',C')$.  Restrict
consideration to the equations labeled by $\gamma'$, corresponding to the
variables $u^{\gamma'}$.  In \eqref{min}, all terms in the second brackets
$[\,\,]$ vanish since all coefficients of $g$ are independent of
$u^{\gamma'}$.  Moreover, the term $g_{\alpha\gamma'}$ vanishes since $g$
has no nontrivial
components $g_{\alpha\gamma'}$, $g_{r\gamma'}$, $g_{A\gamma'}$.  Since
$g_{\alpha'\gamma'}=r^{-2}\bar{g}_{\alpha'\gamma'}$, we conclude that the
$\gamma'$ piece of \eqref{min} reduces to
\be\label{mins}
(\sqrt {\det h})^{-1}\partial_\beta\left [r^{-2} \sqrt{\det h} \,\,
  h^{\alpha\beta}  \partial_\alpha u^{\alpha'}\right ] = 0,
\ee
where $\alpha$, $\beta$ still represent triples $(\alpha,r,A)$,
$(\beta,r,B)$.

According to \eqref{Z0} and \eqref{indmetric}, we have
$$
h=
\begin{pmatrix}
r^{-2}\gb_{\alpha\beta}&0&0\\
0&r^{-2}(1+|v|^2)&r^{-1}\langle v,v_{,B}\rangle\\
0&r^{-1}\langle v_{,A},v\rangle&g_{AB}+\langle v_{,A},v_{,B}\rangle
\end{pmatrix}.
$$
Here $|\cdot|$ and $\langle \cdot,\cdot\rangle$ denote the norm and inner
product on $\mathbb{R}^{n-k}$ determined by $\bar{g}_{\alpha'\beta'}$,
$v_{,A}$ denotes
$\partial_{s^A}v$, $g_{AB}$ is evaluated at $(s,t=0)$, and the blocks
correspond to the decomposition $(\alpha,r,A)$.
Denote by $H$ the $(1+l)\times (1+l)$ matrix (recall $l=\dim \Sigma$):
$$
H=
\begin{pmatrix}
H_{rr}&H_{rB}\\
H_{Ar}&H_{AB}
\end{pmatrix}=
\begin{pmatrix}
1+|v|^2&\langle v,v_{,B}\rangle\\
\langle v_{,A},v\rangle&g_{AB}+\langle v_{,A},v_{,B}\rangle
\end{pmatrix}.
$$
It is evident that $H$ is positive definite, since
$$
H=
\begin{pmatrix}
1&0\\
0&g_{AB}
\end{pmatrix}+
\begin{pmatrix}
|v|^2&\langle v,v_{,B}\rangle\\
\langle v_{,A},v\rangle&\langle v_{,A},v_{,B}\rangle
\end{pmatrix}
$$
is the sum of a positive definite matrix and a positive semidefinite
matrix.  For a given function $v:\Sigma\rightarrow \bbR^{n-k}$, $H$ can be
interpreted as a
metric on $\bbR \times \Sigma$ which is translation-invariant in the $r$
direction.
Now
$\sqrt{\det h} = r^{-k-1}\sqrt{\det \bar{g}_{\alpha\beta}}\sqrt{\det H}$.
Thus \eqref{mins} becomes
\be\label{simpleeq}
(\sqrt{\det H})^{-1}\partial_\beta\left [r^{-k-3} \sqrt{\det H} \,\,
  h^{\alpha\beta}  \partial_\alpha u^{\alpha'}\right ] = 0.
\ee
Since $u^{\alpha'}=rv^{\alpha'}(s)$, only the $(r,A)$ pieces of $\alpha$
enter into the summation.
The left-hand side of \eqref{simpleeq} is of the form $r^{-k-2}$ times a
function of $s$ alone.  This function of $s$ can be identified by expanding
the $\alpha$, $\beta$ sums.  One finds that \eqref{simpleeq} is equivalent
to:
\be\label{Seq}
(\sqrt{\det H})^{-1}\partial_B \left[\sqrt{\det H}
\left(H^{AB}  v^{\alpha'}_{,A} +
H^{rB}  v^{\alpha'}\right)\right]
-(k+1) \left(H^{Ar}  v^{\alpha'}_{,A}
+ H^{rr}v^{\alpha'}\right)=0,
\ee
where we write
$$
H^{-1}=
\begin{pmatrix}
H^{rr}&H^{rB}\\
H^{Ar}&H^{AB}
\end{pmatrix}.
$$
Equation \eqref{Seq} exhibits concretely
the partial differential equation on $v^{\alpha'}(s)$ implied by the
condition that $Z^0$ is minimal for $g^0+g_K$.

We now show that $v^{\alpha'}=0$ is the only global solution of \eqref{Seq} on
a compact manifold $\Sigma$ with metric $g_{AB}$.  Since
$\sqrt{\det H}$ defines a volume form on $\bbR\times \Sigma$ which is
independent of $r$, it can also be interpreted as a volume form on
$\Sigma$ itself.  Suppose $v^{\alpha'}$ solves \eqref{Seq}.  Multiply 
\eqref{Seq} by $\bar{g}_{\alpha'\beta'}v^{\beta'}\sqrt{\det H}$ and
integrate by parts over $\Sigma$.  This gives
$$
\int_\Sigma \left[H^{AB}\langle v_{,A},v_{,B}\rangle
+(k+2)H^{Ar}\langle v_{,A},v\rangle
+(k+1)H^{rr}|v|^2\right]\,\sqrt{\det H}\, ds=0 .
$$
Decompose the integrand as
$$
\left(H^{AB}\langle v_{,A},v_{,B}\rangle
+2H^{Ar}\langle v_{,A},v\rangle
+H^{rr}\langle v,v\rangle\right)
+k\left(H^{Ar}\langle v_{,A},v\rangle
+H^{rr}|v|^2\right).
$$
Since $H^{-1}$ is positive definite, the first term is nonnegative.
So if we can show that
\be\label{nonnegative}
H^{rr}|v|^2+H^{Ar}\langle v_{,A},v\rangle\geq 0,
\ee
then both terms separately must vanish.  {From} the vanishing of the
first term we can conclude that $v^{\alpha'}=0$ as desired.

It remains to establish \eqref{nonnegative}.  Let
$$
\left (
\begin{array}{ll}
C^{rr} & C^{rB} \\
C^{Ar} & C^{AB}
\end{array} \right )
$$
denote the cofactor matrix of $H$, so that
$$
\det H = H_{rr}C^{rr}+H_{Ar}C^{Ar}
$$
and
$$
\left (
\begin{array}{ll}
H^{rr} & H^{rB} \\
H^{Ar} & H^{AB}
\end{array} \right )
=(\det H)^{-1}
\left (
\begin{array}{ll}
C^{rr} & C^{rB} \\
C^{Ar} & C^{AB}
\end{array} \right ).
$$
In particular,
$H^{rr}|v|^2+H^{Ar}\langle v_{,A},v\rangle=(\det H)^{-1}
(C^{rr}|v|^2+C^{Ar}\langle v_{,A},v\rangle)$.
But the cofactor expansion along the first column gives
$$
C^{rr}|v|^2+C^{Ar}\langle v_{,A},v\rangle
=
\det
\left (
\begin{array}{ll}
|v|^2 & H_{rB} \\
\langle v_{,A},v\rangle & H_{AB}
\end{array} \right )
=
\det
\left (
\begin{array}{ll}
|v|^2 & \langle v,v_{,B}\rangle \\
\langle v_{,A},v\rangle & g_{AB}+\langle v_{,A},v_{,B}\rangle
\end{array} \right ).
$$
The matrix
$$
\left (
\begin{array}{ll}
|v|^2 & \langle v,v_{,B}\rangle \\
\langle v_{,A},v\rangle & g_{AB}+\langle v_{,A},v_{,B}\rangle
\end{array} \right )
$$
is positive semidefinite for the same reason that $H$ was positive
definite:  it can be written as the sum
$$
\left (
\begin{array}{ll}
0 & 0 \\
0 & g_{AB}
\end{array} \right )
+
\left (
\begin{array}{ll}
|v|^2 & \langle v,v_{,B}\rangle \\
\langle v_{,A},v\rangle & \langle v_{,A},v_{,B}\rangle
\end{array} \right )
$$
of two positive semidefinite matrices.  Therefore its determinant is
nonnegative.  This establishes \eqref{nonnegative} and so concludes the
argument.

We remark that the vanishing of $v^{\alpha'}$ can alternately be proved by
an integration by parts argument on $\bbR \times \Sigma$ rather than on
$\Sigma$.  Namely, multiply \eqref{simpleeq} by
$g_{\alpha'\beta'}u^{\beta'}\sqrt{\det H}\,ds\,dr$,
integrate $r$ over $(a,b)$ and $s$ over $\Sigma$, where $0<a<b<\infty$ are
fixed, and then integrate by parts in both $r$ and $s$.  The $r$
integration gives rise to a boundary term, but \eqref{nonnegative} implies
that it has a sign.

For analytic metrics $g_{AB}$, nonzero local solutions $v^{\alpha'}$ of
\eqref{Seq} can be constructed as convergent power series
(Cauchy-Kowalewski Theorem).  By appropriately choosing the background
metric $g_K$, it can be arranged that the submanifold $Z^0$ given by
\eqref{Z0} for such a solution $v^{\alpha'}$ is minimal with respect to
$g^0 + g_K$ while at the same time $\Sigma$ is not minimal with respect to
$g_K$.  Such a $Z^0$ thus provides a local example of a minimal submanifold
of a product space $X\times K$ whose boundary is a product $N\times \Sigma$
with $\Sigma$ not minimal with respect to $g_K$.

\section{Examples}\label{examples}

\subsection{Slipping modes on spheres}

\subsubsection{The flavor D7 brane}
\label{flavord7}

Let us begin by reviewing one example in which a ``$t$'' variable is a
non-trivial function of
$r$ as first presented in \cite{Karch:2002sh}. The submanifold fills all of
$X$, which for simplicity we take to be
(Euclidean) $AdS_5$ parametrized in standard Poincar\'e coordinates
$$
ds^2_{AdS} = \frac{1}{r^2} (dr^2 + d \vec{x}^2 ).
$$
$\vec{x}$ are Cartesian coordinates along the $\mathbb{R}^4$ factor.
Let us take the internal space to be $S^5$ with metric written as
$$
ds^2 = d\theta^2 + \cos^2 (\theta) d \Omega_3^2 + \sin^2 (\theta) d\psi^2
$$
where $d\Omega_3^2$ is the round metric on the 3-sphere.

We are looking for a
submanifold of the form
$$
\psi=const., \,\,\,\,\, \theta=\theta(r)
$$
with $\theta(0)=0$.
This defines an 8-dimensional submanifold of the 10 dimensional product
space. Asymptotically, the submanifold fills all of $AdS_5$ ($k=n=4$) and
it wraps an equatorial $S^3$ inside $S^5$ ($l=3$). This submanifold is the
worldvolume of a D7 brane, where the standard physics nomenclature defines
a D$p$ brane as an object extended in $p$ spatial dimensions and hence with
a $p+1$ dimensional worldvolume.

The ansatz describes a submanifold where the $S^3$ wrapped by the D7
shrinks as a function of the radial coordinate:  the D7 ``slips
off". Correspondingly, $\theta$ is often referred to as the ``slipping mode"
in the physics literature.
If $\theta(r)$ reaches $\pi/2$ at any finite $r$, the internal sphere
shrinks to zero size at that point. For a generic submanifold of this type
there will be a singularity at this point. Imposing regularity as an
additional constraint results in a unique submanifold for
given local data on the boundary.

To find the solution, one starts with the area functional restricted to this ansatz:
$$
\label{sevenaction}
{\cal A} = r^{-5} \cos^3 (\theta) \sqrt{1+r^2 (\theta')^2}
$$
and treats it as the Lagrangian of a classical mechanics problem. The
resulting Euler-Lagrange equation gives a non-linear 2nd order ordinary
differential equation for $\theta(r)$.
Nevertheless, it is easy to verify that
\be\label{arcsin}
 \theta = \arcsin (m r),
\ee
 where $m$ is a constant, defines a one parameter family of
 solutions\footnote{The reason that such a simple solution exists is
   supersymmetry. Instead of solving the 2nd order differential equations
   directly one can find this $\arcsin$ solution by solving an auxiliary
   problem of finding a particular Killing spinor, which amounts to solving
   a first order equation.  See \cite{Karch:2002sh}.}.
For this solution, we can expand near the boundary
$$
 \theta \sim mr - \frac{(mr)^3}{6} + {\cal O}(r^5).
$$
The local data is the coefficient $m$ of $r^{\Delta_-} = r$. The global
data is the coefficient $-m^3/6$ of
$r^{\Delta_+}=r^3$; for this solution it is indeed fixed in terms of the
local data.
The submanifold does not extend past $r=1/m$. At $r=1/m$ the submanifold
terminates smoothly. The 3-sphere it is wrapping inside the 5-sphere
shrinks to zero size, but locally the induced metric just becomes flat
$\mathbb{R}^8$ and there is no curvature singularity.

The physical interpretation of this flavor brane is as outlined in the
introduction. It adds fundamental representation matter (in this case a
hypermultiplet preserving 8 supercharges) to the field theory which
otherwise only hosts adjoint representation fields (${\cal N}=4$ SYM in
this case). The ``slipping mode" $\theta(r)$ in the field theory maps to a
bi-linear operator made from two defect fields. Turning on the coefficient
of $r^{\Delta_-}=r$ corresponds in the field theory to adding a mass for
the flavors.  The coefficient of $r^{\Delta_+}=r^3$ is related to the
vacuum expectation value of this bi-linear operator, the ``chiral
condensate"; the precise relation has been worked out e.g. in
\cite{Karch:2005ms} using the technique of holographic renormalization
(holo-RG). The regularity condition in the interior (the IR
boundary condition) fixes the relation between the two; the $\arcsin$
solution is determined by a single parameter $m$.

\subsubsection{General AdS times sphere example}\label{spheres}

The example of the previous subsection can easily be generalized to
submanifolds that asymptote to AdS$_{k+1}$ $\times$ $S^l$ inside
AdS$_{n+1}$ $\times$ $S^m$.
Now we write
$$
ds^2_{AdS} = \frac{1}{r^2} (dr^2 + d \vec{x}^2 + d \vec{y}^2)
$$
with $\vec{x}\in \mathbb{R}^k$ and $\vec{y}\in \mathbb{R}^{n-k}$.
As above, write the metric on
$S^m$ as
\be\label{spheremetric}
ds^2 = d \theta^2 + \cos^2(\theta) d \Omega_l^2 + \sin^2(\theta) d \Omega_{m-l-1}^2.
\ee
To keep the discussion
general, we also allow for a slightly more general background metric: we
study submanifolds in AdS$_{n+1}$ $\times$ $S^m$ where we allow the radius
of curvature $R_s$ of the internal sphere $S^m$ to be different from the
radius of curvature $R=1$ of AdS$_{n+1}$, with $(R_s/R)^2 = \alpha$.  
The $\vec{y}$ variables and the variables in $S^{m-l-1}$ are constant on
the submanifolds under consideration so these variables will play no role.

One possible minimal area submanifold in all these cases is the trivial
$(k+1+l)$-dimensional submanifold $\theta=0$,
which is globally AdS$_{k+1}$ $\times$  $S^l$, not just asymptotically
close to the boundary.
The Lagrangian (the area element) for the slipping mode $\theta(r)$ is
$$
{\cal A} = r^{-k-1} \cos^l (\theta) \sqrt{1+\alpha r^2 (\theta')^2}.
$$
While $\theta=0$ is a solution, we can deform the cycle by turning on
$\theta$ and only requiring that
$\theta$ go to zero asymptotically. Close to the boundary we can linearize
the equations in $\theta$.
The slipping mode acts like a scalar field with mass squared $M^2=-l/\alpha$ in
AdS$_{k+1}$. Its near boundary behavior is $r^{\Delta}$ with the usual
$$
\Delta_{\pm} = \frac{k}{2} \pm \frac{1}{2} \sqrt{k^2 + 4 M^2}
 = \frac{k}{2} \pm \frac{1}{2} \sqrt{k^2 - 4 l/\alpha}.
$$
$\Delta$ becomes complex for $\alpha<\alpha_{crit} =4l/k^2$.

For the D7 example above $\alpha=1$, $k=4$ and $l=3$ so $\Delta=1$ and $\Delta=3$ are
the two solutions. A similarly nice
example \cite{Karch:2000gx} is the D5 brane with AdS$_4$ $\times$ $S^2$
asymptotics in AdS$_5$
$\times$ $S^5$.  Then $\alpha=1$, $k=3$ and $l=2$, so that $\Delta=1$ and
$\Delta=2$ are the two solutions. In this case the arcsine of
\eqref{arcsin} is once more an analytic solution. In fact, it is easy to
check that \eqref{arcsin} solves the equations of motion whenever
$\alpha=1$ and $k=l+1$. The corresponding dimensions $\Delta$ again turn
out to be integers in this case, $\Delta=1$ and $\Delta=k-1$.
But in general $\Delta$ is irrational.
If one in addition takes the slipping mode to depend on the $S^l$
coordinates, one finds solutions (to the linearized equations) where
$\theta$ is a spherical harmonic (eigenfunction of the Laplacian with
eigenvalue $L (L+l-1)$) on the internal space and its effective mass
squared in AdS is then
\be\label{masssquared}
\alpha M^2 = -l + L (L+l-1).
\ee
For $\alpha =1$ and $l=k-1$ this gives $\Delta=1-L$ and $\Delta=k-1+L$, and 
in particular for the D7 example one has $\Delta=1-L$ and $\Delta = 3+L$. 

The two different $\Delta$'s have standard interpretation in the physics
literature. The smaller $\Delta$ corresponds to a non-normalizable
mode. Turning it on amounts to deforming the theory. This way one naturally
obtains a one-parameter family of submanifolds. This is exactly the parameter
$m$ in the D7 example before.
Only if the mass squared lies between $-k^2/4$ and 0 does one get a
positive $\Delta$ for this non-normalizable mode.
The larger value of $\Delta$ corresponds to the normalizable mode. As in the D7 example,
its coefficient is usually fixed by a regularity condition in the interior. If
the mass squared is positive we get one positive and one negative
$\Delta$. So if we want the submanifold to be regular at $r=0$, in this
case one is limited to the larger (normalizable) $\Delta$. In the physics 
language the positive mass squared corresponds to ``irrelevant"
operators. When added to the field theory lagrangian (= turn on the
leading behavior of the scalar) they grow at high energies (=close to the
boundary); their backreaction destroys the AdS asymptotics and so we do not
consider them here.
They can still have a non-trivial expectation value (=
coefficient of the subleading term) which is determined dynamically (=by a
regularity condition in the interior).

In terms of physics, the interpretation of the more general flavor branes
discussed in this subsection
is very similar to the D7 example above. Some fundamental representation
matter is added to the field theory. Of course for $\alpha \neq 1$ and
general $k$ and $l$, we don't have a known duality that realizes this
background/minimal area pair. If $k<n$,
the flavor is localized on a defect, for $k=n$ it is spacetime-filling as
in the D7 case. In all these cases, one expects a single regular solution to
exist for a given coefficient of the $r^{\Delta_-}$ asymptotic term.
The simple analytic solution in terms of an $\arcsin$ is explained by
supersymmetry for the D7 and the D5 examples above.

In \eqref{geometricJ} we gave a general formula for the Jacobi
operator. Let us confirm that this reproduces the same results for these
examples.
An equatorial $S^l\subset S^m$ is totally geodesic,
so $\mathcal{F}=0$, and one sees easily that $\mathcal{R}=l\,\text{Id}$
(unless $m=1$, in which case $\mathcal{R}=0$ even if $l=1$).  The
normal bundle is trivial with a global parallel frame.  So the spectrum of
$\nabla^*\nabla$ on $\mathcal{N}$ consists of the spectrum of the scalar
Laplacian on $S^l$, namely $\{L(L+l-1)\}$, with multiplicity $m-l$. For $\alpha=1$ this
gives $\lambda= L(L+l-1)-l$ and is consistent with the discussion above.
The rescaling observation in \S\ref{intfluc} produces the correct
dependence on $\alpha$ in \eqref{masssquared}.

Note that $\lambda
=0$ for $L=1$, so there is exactly one negative $\lambda$, and that this
persists under general rescaling of $g_K$.  Thus the choice $L=0$ is the
only possibility to obtain $\Re\Delta_->0$, i.e. we can only have one
relevant deformation for any $S^l\subset S^m$ and any scaling of
$g_K$.

\subsection{Disjoint boundaries}

\subsubsection{Wilson lines with internal motion}
An example with both ``$t$'' and ``$u$'' variables being non-trivial
functions of
$r$ is the Wilson line of \cite{Maldacena:1998im}. In this case the boundary
is disconnected with two components: $\partial Z =
(N_1 \times \Sigma_1) \cup (N_2 \times \Sigma_2 )$.  Generalizations of
this example will allow us to
study in detail the backreaction of the ``$u$''-variables on the
``$t$''-variables
and vice versa. For the case of the rectangular Wilson line in
AdS$_5$ $\times$ $S^5$ discussed in \cite{Maldacena:1998im}, $N_{1,2}$ are
parallel straight lines separated by
a distance $\Delta u$ and $\Sigma_{1,2}$ are points separated by a distance
$\Delta \theta$ on the internal sphere. Locally, close to each of the
components of the boundary, the minimal area asymptotes to AdS$_2$ times a
point, but globally the two asymptotic regions are connected into one
smooth U-shaped minimal surface.  The special case $\Delta \theta =0$
corresponds simply to a minimal surface in AdS$_5$.  A second example
\cite{Skenderis:2002vf,Karch:2005ms,Evans:2013jma} has a D5 brane with two
locally $AdS_4 \times S^2$ asymptotic regions ending on two parallel
$\mathbb{R}^3 \times S^2$ boundaries, where the $S^2$ in both cases is the
same equatorial $S^2$ inside the $S^5$.  The equatorial $S^2$ does not
move: this example is a product of a
four-dimensional minimal submanifold of AdS$_5$ with $S^2$.
All of these examples fit into the more
general framework of having $\partial Z$ be two copies of
$\mathbb{R}^k \times S^l$ for which the two $\mathbb{R}^k$'s are
parallel. In the next two subsections we consider two
different generalizations of these examples to $\mathbb{R}^k \times S^l$:
\begin{itemize}
\item The most direct generalization of the example of
  \cite{Maldacena:1998im} has two disjoint copies of $\mathbb{R}^k \times
  S^l$
where the background internal space is $S^{2l+1}$ and the two $S^l$'s are
equatorial and disjoint.
They still can be connected by a smooth U shaped submanifold. This set of
examples in particular contains
the Wilson line with separate points on the internal sphere as the special
case of $l=0$.
\item The second generalization has two disjoint copies of $\mathbb{R}^k
  \times S^l$ with the same equatorial $S^l$ for both, as in the D5 example
  above.  But now we turn on a non-trivial slipping mode on each of the
  disjoint defects, leading to an interesting interplay between internal
  and AdS coordinates.
\end{itemize}

The physics interpretation of a single such defect was discussed in the
previous subsection: each defect adds fundamental matter to the gauge
theory, localized on $N_1$ and $N_2$ respectively. For a connected
worldvolume
in the bulk to be allowed, one needs one of the defects to be an
anti-defect. Flavor D-branes come with an orientation, and in the connected
worldvolume the orientation changes between the two defects. This was
already the case in the Wilson line example, where the U-shaped worldvolume
evaluates the quark/anti-quark potential, not the quark-quark
potential. For the latter only a disconnected worldvolume is allowed in the
bulk. For quark/anti-quark both connected and disconnected configurations
are allowed (and often there is a competition between the two for which one
has the lower area). Such brane/anti-brane configurations typically break
all supersymmetry. In the D5 example, both D5 and anti-D5 individually
preserve half the supersymmetry, but it is the opposite half that is
preserved. Together they break supersymmetry completely.

\subsubsection{Rotating spheres}\label{rotate}

Let us first look at a case of two disjoint copies of $\mathbb{R}^k
\times S^l$, where the background internal space is $S^{2l+1}$ and the two
$S^l$'s are equatorial and disjoint.  We write
\be\label{adsmetric}
ds^2_{AdS} = \frac{1}{r^2} (dr^2 + d \vec{x}^2 +du^2 + d\vec{y}^2)
\ee
with $\vec{x}\in \mathbb{R}^k$, $u\in \mathbb{R}$, $\vec{y}\in
\mathbb{R}^{n-k-1}$.  Once again $\vec{y}$ plays no role.
We are looking for a maximally
symmetric solution which is translation-invariant in $\vec{x}$ and where
$u(r)$ is turned on as the only ``$u$"
variable. Asymptotically, $u(r)$ should be a double-valued function which
approaches $u(0) = \pm \Delta u/2$. The two branches will be smoothly
connected at a turning point at $r_{max}$ with $u(r)\sim \sqrt{r_{max}-r}$
close to $r_{max}$.

To get the Lagrangian, we write the $S^{2l+1}$ metric as
$$
ds^2 = d \theta^2 + \cos^2(\theta) \, dS_l^2 + \sin^2(\theta) \,
d\tilde{S}_l^2. 
$$
We embed $S^l \times S^1$ into $S^{2l+1}\subset \mathbb{R}^{2l+2}$ by
$$
(w,\theta) \rightarrow (\cos(\theta) \, w, \sin(\theta) \, w) 
$$
where $w\in \mathbb{R}^{l+1}$, $|w|=1$.  Then the
$S^{2l+1}$ metric pulls back to just
$$
ds^2= d \theta^2 + dS_l^2(w).
$$
Adding this to the usual AdS metric and then taking $u$ and $\theta$ to be
functions of $r$ as before, the $dS_l^2(w)$ does not interact with anything
in forming the Lagrangian, and the effective Lagrangian is independent of
$l$:
$$
{\cal A} =r^{-k-1} \sqrt{1 + (u')^2 +  \alpha r^2 (\theta')^2}.
$$
As a result, the submanifold for {\it any} $l$ is just given by the
generalization of the solution in \cite{Maldacena:1998im} from $k=1$ and
$\alpha =1$ to arbitrary $k$ and $\alpha$, but independent of $l$.  This
solution is derived by observing that $u$ and $\theta$ do not appear in
${\cal A}$, so there are two conserved quantities.  This leads to
\be\label{xmaldacena}
u' = \pm \, c_1 \frac{r^{k+1}}{\sqrt{ 1 - c_1^2
    r^{2k+2}  -  \alpha c_2^2 r^{2k}}}, \quad
\quad \theta'= \frac{c_2}{c_1r^2}\, u' ,
\ee
from which $u$ and $\theta$ are obtained by integration.
The two integration constants $c_1$ and $c_2$ set the separations
$\Delta u$ and $\Delta \theta$ of the two disjoint boundary pieces.  The
case $c_2=0$ corresponds to a product solution with a U-shaped
minimal submanifold of hyperbolic space.  The
remaining two integration constants set the overall position. They can
always be set to zero by exploiting translation invariance in $u$ and
$\theta$.

Let us briefly see how these exact solutions of the full non-linear system
fit into our general description.
As we noted at the end of \S\ref{spheres}, we get that $\alpha M^2
=0$ for $L=1$ for any dimension $l$ of
the internal sphere (and for any $\alpha$). By locking the two $S^l$ spheres
to each other, we implicitly turned on an $L=1$ mode. With this, we get
$\Delta_-=0$ and $\Delta_+=k$ for the $\theta$ mode, which is consistent
with the solution \eqref{xmaldacena}.

From the physics point of view it is somewhat surprising why the $L=1$ mode
is so special in the sense that a simple solution to the full non-linear
equations can be found for any $l$. Typically the slipping mode describes a
bi-linear operator made out of two defect fields, whereas $L=1$ describes a
tri-linear operator made of two defect fields and one of the
adjoint fields. The dual solution has the brane rotating inside the
internal $S^{2l +1}$ along the U-shaped worldvolume.

\subsubsection{Interaction between internal and hyperbolic factors}
For $l \geq 1$ we can consider a second class of examples, where this time
the two disjoint asymptotic defects wrap
the same $S^l$, but we turn on a slipping mode. As before,
asymptotically the slipping mode will scale\footnote{For $l=0$ one has
  $\Delta_-=0$ so turning on
the non-normalizable piece of the slipping mode has in this case the
interpretation of actually separating the points and so is identical to the
case discussed in the previous subsection.} as $\theta \sim
r^{\Delta_-}$. The coefficient of this $r^{\Delta_-}$ does not have
to be the same on the two disjoint defects.

We write the AdS metric in the form \eqref{adsmetric} and the
$S^m$ metric in the form (\ref{spheremetric}), with an overall prefactor
of $\alpha$ for the metric on $S^m$ to account for
the difference in curvature radii.  The $\vec{y}$ variables and the
variables in $S^{m-l-1}$ are held constant.  We are interested in
solutions where we turn on the slipping mode $\theta(r)$ as our only
``$t$'' variable, and $u(r)$ as our only ``$u$" variable.
This ansatz is forced upon us if we insist on preserving
the full symmetry of $\mathbb{R}^k \times S^l$ as well as the isometries of
the transverse $S^{m-l-1}$.   With this ansatz, the Lagrangian for the area
of the submanifold parametrized by $\theta(r)$ and $u(r)$ is proportional
to
$$
{\cal A} = \frac{\cos^l (\theta)}{r^{k+1}} \sqrt{1 + (u')^2 + \alpha r^2 (\theta')^2}.
$$
As $u$ only appears derivatively (as a consequence of translation
invariance of the background metric) we can solve for $u(r)$ explicitly
using an integral of motion:
$$
c = \frac{\delta {\cal A}}{\delta u'} = \frac{\cos^l (\theta)}{r^{k+1}}
\frac{u'}{\sqrt{ 1 + (u')^2 + \alpha r^2 (\theta')^2}}
$$
which is easily solved for $u'$
\be\label{xsolution}
u' = \pm \, c r^{k+1} \frac{\sqrt{1 + \alpha r^2
    (\theta')^2}}{\sqrt{\cos^{2l} (\theta) - c^2 r^{2k+2}}}. \ee
The equations are quadratic in $u'$, so we get a free sign choice. The two
allowed choices correspond to the two branches. To obtain the equations of
motion of $\theta(r)$ we want to substitute $u'(r)$ back into the original
Lagrangian. One has to be careful though that, while the original
Lagrangian instructed us to vary with respect to $\theta$ at fixed $u$ and
$u'$, we now want to keep $c =  \delta {\cal A}/\delta u'$ fixed, that
is the conjugate momentum. The correct action from which to derive the
equation of motion for $\theta(r)$ by varying with respect to $\theta$ at
fixed $c$ is the Legendre transformed
\be
\label{thetaaction}
\tilde{\cal A}= {\cal A} - u' \, \frac{\delta {\cal A}}{\delta u'} = \frac{1}{r^{k+1}}
\sqrt{1+ \alpha r^2 (\theta')^2} \sqrt{\cos^{2l} (\theta) - c^2 r^{2k
    +2}}.
\ee
In the special case $l=0$ this $\tilde{\cal A}$ only depends on $\theta'$,
not on $\theta$, and the full system can be solved analytically as in
\S\ref{rotate}.
For general $l$ one has to resort to numerics to
construct $\theta(r)$, as has recently been carried out in
\cite{Chang:2013toa} for $k=3$, $l=2$.

For this family of examples we can study higher order terms in
the near boundary expansion analytically. All we need to do is inspect our
explicit solution for $u'$ as
well as the form of the effective action for $\theta$. Let us first look at
the asymptotic form of $\theta(r)$. Close to the boundary $\theta(r)$
vanishes and we can determine its behavior by expanding $\tilde{\cal A}$ to
quadratic order in $\theta$ as well as in $r$. Dropping irrelevant $\theta$
independent terms we get
\be\label{quadraticaction}
\tilde{\cal A} \sim \frac{1}{2 r^{k+1}}\left ( \alpha r^2 (\theta')^2 (1 -
\frac{1}{2} c^2 r^{2k+2} + \ldots)  - l \theta^2 \right ).
\ee
To find the leading near boundary behavior we can neglect the subleading
$c^2 r^{2k+2}$ term, which gives us the backreaction of $u$ on
$\theta$. The remaining action is just the one of a scalar field of mass
squared $M^2=
-l/\alpha$ in AdS$_{k+1}$. Correspondingly, the two possible boundary
behaviors are once more
$$
\Delta_{\pm} = \frac{k}{2} \pm \frac{1}{2} \sqrt{k^2 - 4 l/\alpha}.
$$
Armed with our knowledge about $\theta(r)$ we can inspect formula
(\ref{xsolution}) for $u'(r)$ to determine the leading near boundary
behavior of
$u$. Since $\theta$ goes to zero at the boundary, the leading small $r$
behavior of $u'$ is given by $u'\sim c r^{k+1}$ and so $c$ represents the
locally undetermined coefficient in the expansion of $u$. It affects $u$ at
order $r^{k+2}$ as expected. $c$ is determined in terms of $\Delta u$ for
the connected configuration. The leading correction due to the backreaction
of $\theta$ comes from the $c\alpha r^{k+3}(\theta')^2$ and
$c r^{k+1}\theta^2$ corrections to $u'$ which
arise from expanding out the square roots and the $\cos^{2l}(\theta)$
term.  This affects $u$ itself at order $r^{k+2+2 \Delta_-}$.  (We are
thinking of the situation $k^2>4l/\alpha$ so that $\Delta_\pm$ are real
with $\Delta_-<\Delta_+$.)  Last but not least, we need to understand
the backreaction of $u$ on
$\theta$. This backreaction is determined by the term proportional to
$c^2$ in the action \eqref{thetaaction} for the $\theta$ fluctuations.
Again droppping $\theta$ independent terms, we can write
\eqref{thetaaction} as
$$
\tilde{\cal A}\sim r^{-k-1}\Big[
\cos^{l}(\theta)\sqrt{1+ \alpha r^2 (\theta')^2}
 -\frac14 c^2 \alpha r^{2k +4}(\theta')^2 +\ldots\Big].
$$
The first term in $[\cdot]$ gives rise to the equation of motion for a
pure slipping mode $\theta_0(r)$ with no $u$ dependence discussed in
\S\ref{spheres}.  If we make an ansatz that $\theta=\theta_0 + \delta
\theta$ with $\theta_0\sim r^{\Delta_-}$, the correction term in the
equation of motion demands that $\delta \theta$ is of order
$r^{2k+2 + \Delta_-}$.  Of course, the global regularity condition in the
interior for the coupled problem will affect the $r^{\Delta_+}$ coefficient
in $\theta$ as well, and this term appears earlier in the expansion than
the $r^{2k+2 + \Delta_-}$ term.

It is straightforward to integrate the equations of motion numerically,
once the smoothness condition in the IR is properly implemented. Explicit
examples and a full phase diagram of these configurations, in particular
addressing the question whether the connected or the disconnected
configuration has the smaller area, have been presented in
\cite{Chang:2013toa}. One interesting new phenomenon that occurs in these
examples is that for a certain range of local boundary data more than one
connected regular minimal area exists, that is the global terms aren't
unique but can be chosen from a discrete family.

\begin{acknowledgments}
We'd like to thank Michael Spillane for collaboration during initial stages
of this work. We'd also like to thank Han-Chih Chang, Carlos Hoyos, Kristan
Jensen, Rafe Mazzeo, and Andrew O'Bannon for useful discussions.
The work of A.~Karch was supported in part by U.S. DOE grant
No. DE-FG02-96ER40956.  The work of R. Graham was partially supported by
NSF grant \# DMS 1308266.
\end{acknowledgments}

\bibliography{physicspaper}
\bibliographystyle{JHEP}

\end{document}